\documentclass{cupconf}
\usepackage{epsfig}

  \checkfont{eurm10}
  \iffontfound
    \IfFileExists{upmath.sty}
      {\typeout{^^JFound AMS Euler Roman fonts on the system,
                   using the 'upmath' package.^^J}%
       \usepackage{upmath}}
      {\typeout{^^JFound AMS Euler Roman fonts on the system, but you
                   dont seem to have the}%
       \typeout{'upmath' package installed. cupconf.cls can take advantage
                 of these fonts,^^Jif you use 'upmath' package.^^J}%
      }
  \else
  \fi

  \checkfont{msam10}
  \iffontfound
    \IfFileExists{amssymb.sty}
      {\typeout{^^JFound AMS Symbol fonts on the system, using the
                'amssymb' package.^^J}%
       \usepackage{amssymb}%

      }{}
  \fi

  \IfFileExists{amsbsy.sty}
    {\typeout{^^JFound the 'amsbsy' package on the system, using it.^^J}%
     \usepackage{amsbsy}}
    {\providecommand\boldsymbol[1]{\mbox{\boldmath $##1$}}}

\newsavebox{\astrutbox}
\sbox{\astrutbox}{\rule[-5pt]{0pt}{20pt}}

\title[r-process, $\alpha$-elements and kinematics in the Galactic disk]{{\it \small \flushleft STScI May Symposium, 3--6 May, 2004, Baltimore, MD, USA\\Planets To Cosmology:  Essential Science In Hubble's Final 
Years\\
M. Livio (ed.)\\\mbox{ }\\} R-process and $\alpha$-elements in the Galactic disk: Kinematic correlations}

\author[A. Koch \& E.K. Grebel]%
{A\ls N\ls D\ls R\ls E\ls A\ls S\ns K\ls O\ls C\ls H\ns \&\ns
E\ls V\ls A\ns K.\ns G\ls R\ls E\ls B\ls E\ls L}

\affiliation{Astronomical Institute of the University of Basel, Venusstr. 7, CH-4102 Binningen, Switzerland}

\begin{document}

\maketitle

\section{Past}
How do disk galaxies form and evolve? What are the detailed evolutionary histories of their various subcomponents, when and how rapidly did they form, and how did they get enriched? 
One possibility of investigating these questions is to conduct Galactic archeology by collecting data on elemental abundance ratios, kinematics and ages of stars in our Milky Way, 
which we can observe at the greatest possible level of detail. The resulting observational findings can then be compared to the predictions of careful chemodynamical modelling.
Recent studies of elemental abundances in the Galactic halo (Gratton et al. 2003) and in the Galactic disk (Bensby, Feltzing \& Nordstr\"om 2003, 2004a; Allende Prieto et al. 2004, Borkova \& Marsakov 2004, Mishenina et al. 2004) 
have underscored the possibility to kinematically separate different Galactic subcomponents. Correlations between the galactocentric rotation velocity and various element ratios were 
found, providing an important means to link different tracers of star formation and metal enrichment to the Galactic components of different origin (collapse vs. accretion). 
In the present work we concentrate on three main tracers: (i) Europium as an r-process element is predominantly produced in Supernovae of type II. (ii) Likewise, $\alpha$-elements, 
such as Ca, Si, Mg, are synthesised in SNe II, contrary to iron, which is being produced preferentially in SNe Ia. (iii) The s-process element Barium is a measure of the relative 
contribution of AGB stars to the Galaxy's enrichment history and has been shown to be an indicator for distinguishing between thin and thick disk stars.
All such studies reveal, 
basically, that stars with low galactocentric rotational velocity tend to have high abundances of $\alpha$-elements and Eu, but lower abundances of, e.g., Ba.

\section{Present}

\subsection{Our samples}
For this work we adopted elemental abundances of the $\alpha$-elements and for the s-process element Barium from the sample of Edvardsson et al. (1993, hereafter E93). Abundance ratios 
for the r-element Europium were drawn from Koch \& Edvardsson (2002), which provides a subsample of the E93 work. The kinematic data of the 189 stars shown here (124 with published 
[Eu/Fe]) were taken from the HIPPARCOS catalogue, which provides accurate positions, parallaxes and proper motions. These were not yet available for E93's purposes. 
Radial velocities are those from the compilations of Barbier-Brossat \& Figon (2000) and Duflot (1995), both accessed via VizieR at the CDS.

\subsection{Our calculations}
From the basic kinematic data stellar orbits were calculated 5 Gyr backwards in the Galactic three-component potential by Allen \& Santill\'an (1991). 
The solar motion was adopted as (U,V,W)$_{\odot}$ = (10, 5.2, 7.2) km\,s$^{-1}$ (e.g., Binney \& Tremaine 1994). Ending up with positions (X,Y,Z) and the stellar velocity 
components, we display in the following plots the galactocentric rotation velocity, defined by the tangential component $\Theta$, and the orbital eccentricity $e$ as derived 
from the apo- and pericentric distances, respectively.
Since E93's data suggests that the [$\alpha$/Fe] ratio is dependent on galactocentric radius, pointing to a more 
rapid star formation in regions with distances smaller than 7 kpc, we chose to divide our sample in stars at radii larger and smaller than this value.

\subsection{Our results}
Figure 1 shows our results for galactocentric rotation velocity versus metallicity for our sample stars compared to the curve from the chemodynamical models of Samland \& Gerhard (2003).
Although our data only cover the metal-rich regime, we find a qualitatively good agreement with the models in this part.

\begin{figure}
\begin{center}

\epsfig{file=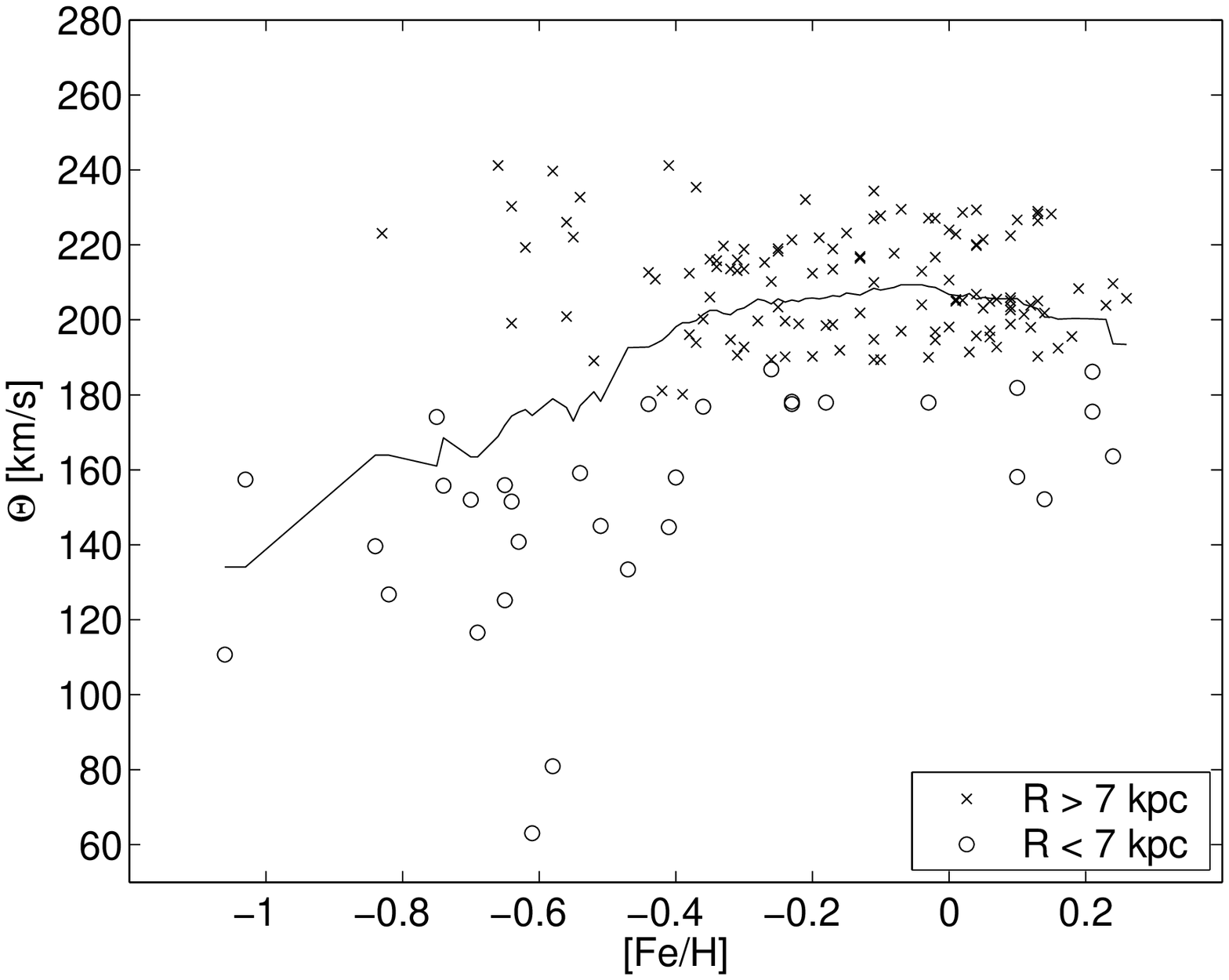, width=8.3cm}

\vspace{0.5cm}

\epsfig{file=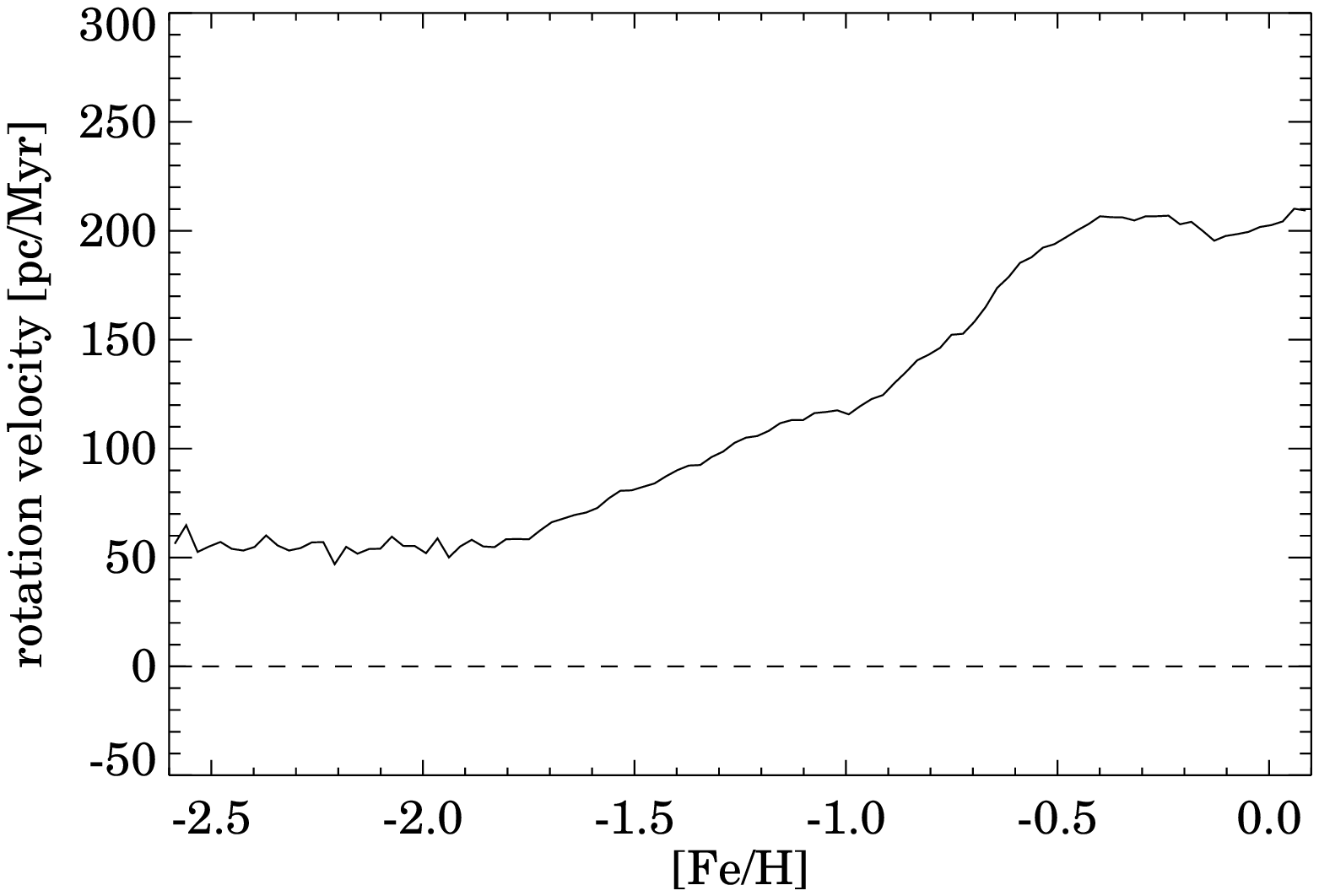, width=8.8cm}
\end{center}
\caption{\textit{Top panel}: Rotation velocity vs. iron abundance for the 189 stars of our sample. The solid line represents the average of our data points. This average is
qualitatively in
agreement with the expected trend from 
chemodynamical models (\textit{bottom panel} - from Samland \& Gerhard 2003).}
\end{figure}

\newpage
\mbox{ }

\vspace{0.2cm}
As Fig. 2a shows, there is no direct linear correlation between europium abundance ([Eu/Fe]) and rotational velocity. It turns out that stars with lower velocities 
(i) are more concentrated towards the inner regions of the Milky Way and 
(ii) show a significantly larger scatter in the  europium-to-iron ratios. 
Basically the same holds for the $\alpha$-elements (Fig. 2b): There is a clear tendency of the $\alpha$-poorer stars to show higher rotation velocities and to be 
located preferentially in the outer regions. On the other hand, the scatter in the abundances is larger in the inner parts, where the velocities are predominantly 
smaller. There is a distinct drop in the $\Theta$ vs. [$\alpha$/Fe] curve at  0.17\,dex, where the faster rotating stars have lower abundances. The same trend is seen 
for the orbital eccentricities: the $\alpha$-poor stars have smaller values of $e$ on average (Fig. 2d). Finally, the opposite trend holds for the [Ba/Eu]-ratio, with 
the stars with reduced Ba abundances rotating more slowly on orbits of systematically higher eccentricity (cf. Figs. 2c, d). Also here, a transition occurs at
[Ba/Eu]$\approx-0.3$\,dex.

Iron is predominantly produced in SNe of type Ia, whereas the $\alpha$-elements and Europium are believed to arise from massive stars including SNe II explosions, 
relating them to a fast-evolving population. In this vein, the occurrence of higher scatter in the Eu- and $\alpha$-to-iron ratios in the inner Galactic regions (R\,$<$\,7\,kpc) 
and the tendency towards an enrichment of these elements indicates that these inner parts presumably had a major contribution from SNe II and thus underwent a rapid formation. 
This could point to a dissipative origin of these inner regions of the Milky Way.
The trends that are seen in the outer disk can be explained by a a lower rate of SNe II, 
a higher occurrence of SNe Ia, or presumably by an enhanced loss of metals from these regions. In this respect, these outer regions resemble the patterns of dwarf spheroidal 
galaxies (e.g., Shetrone et al. 2003) and thus can be attributed to an accretion component of the Galactic disk.   

Furthermore, the results of Bensby
\ et al. (2004b) indicate the presence of an age-metallicity relation in the Galactic disk as well as extended stars formation with SNe Ia
contributions peaking for 3--4\,Gyr in the thick disk.

\section{Future}
While more and more data on individual stars in the Milky Way are being collected -- for instance, the Geneva-Kopenhagen survey assembled data for 13500 stars (Nordstr\"om et al.); 
moreover, the 
Radial Velocity Experiment (RAVE, Steinmetz 2003), aims at observing 50 
million stars from all Galactic substructures out to 60 kpc -- nearby and distant disk galaxies offer a range of possibilities for testing and improving galaxy evolution models. 
Here the emphasis is not on detailed element, but on the colour evolution of unresolved stellar populations. Predictions from chemodynamical models are readily compared to high 
signal-to-noise multi-wavelength imaging data of disk galaxies to study their evolution as a function of redshift, a vital complement to the detailed studies in the nearby Universe. 
Here HST with its unparalleled sensitivity, resolution, and wavelength coverage will play a major and unique role in the years to come.

\begin{acknowledgements}
\mbox{ }\\
{\em \small Acknowledgements} {\small We thank Michael Odenkirchen for kindly providing his codes for 
orbit calculations.
AK and EKG gratefully acknowledge support by the Swiss National Science 
Foundation through grant 200021-101924/1.
This work has made extensive use of the VizieR Service, operated at the CDS, Strasbourg, France.}
\end{acknowledgements}

\clearpage

\begin{figure}
\vspace{.7cm}
\hspace{-1.25cm}
\epsfig{file=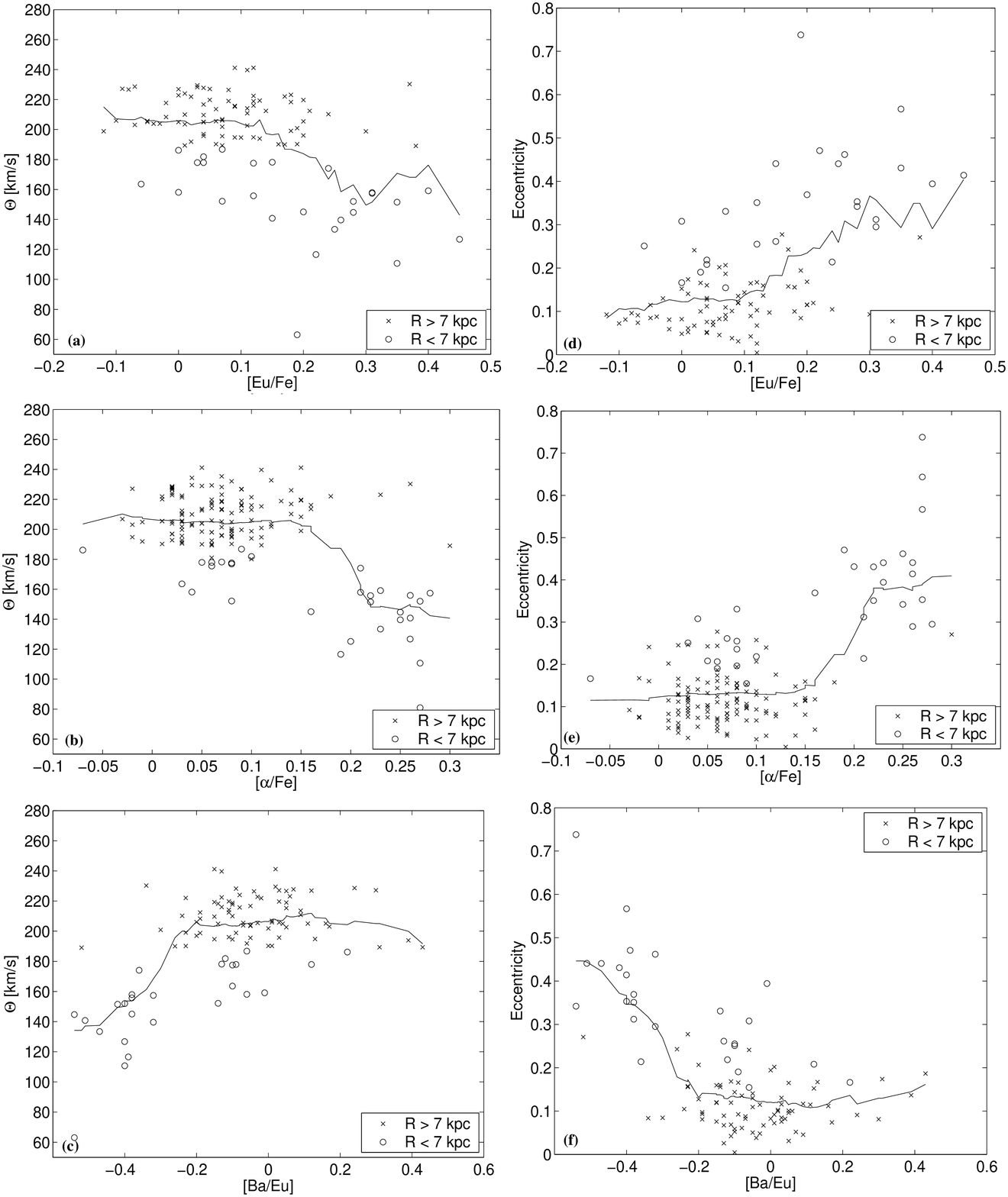, width=16cm}
\vspace{0.5cm}
\caption{Elemental abundances for Eu (\textit{top panels}, $\alpha$-elements (\textit{middle}) and Ba-Eu-ratio (\textit{bottom}) versus kinematic properties: galactocentric rotational
velocity (\textit{left panels}) and orbital eccentricity (\textit{right}). Indiceated as a solid line is the average of the respective data. The data have been separated into two subsets
according to their mean galactocentric radius.}
\end{figure}

\end{document}